# Production d'hydrogène par oxydation catalytique partielle du méthane. Etude du mécanisme réactionnel.

FLEYS Matthieu[a], SIMON Yves[a], MARQUAIRE Paul-Marie[a*], LAPICQUE François[b]

[a]Département de Chimie Physique des réactions, U.M.R. 7630 - CNRS

[b]Laboratoire des Sciences du Génie Chimique, U.P.R. 6811 - CNRS

Nancy Université - ENSIC, 1, rue Grandville – BP 451, 54001 NANCY Cedex

**Résumé**

La réaction d'oxydation partielle du méthane (OPM) a été étudiée expérimentalement en utilisant un réacteur auto-agité catalytique (RAC) sur une large plage de conditions expérimentales. La détermination du mécanisme réactionnel a été effectuée par comparaison entre les résultats expérimentaux et ceux obtenus par simulation grâce au logiciel de simulation de mécanismes réactionnels Chemkin-Surface. Le mécanisme proposé comporte 444 processus homogènes et 33 réactions de surface.

**Mots-clés :** Méthane, oxydation, hydrogène, simulation, $La_2O_3$, mécanisme

## 1. Introduction

Parmis les sources énergétiques disponibles, le gaz naturel fait partie des plus abondantes. Il est majoritairement constitué de méthane dont la conversion directe est très gourmande en énergie du fait de la valeur élevée de l'énergie de la liaison C-H (104 kcal.mol$^{-1}$). La plupart des études de conversion directe du méthane sont donc restées au stade expérimental. Au contraire, la transformation indirecte du méthane en passant par la formation de gaz de synthèse (mélange : $CO + H_2$) est quant à elle beaucoup plus intéressante et viable économiquement. Le vaporéformage est aujourd'hui le procédé le plus répandu dans la production de gaz de synthèse. Cependant, cette réaction est coûteuse à mettre en œuvre car elle est fortement endothermique. La réaction d'oxydation catalytique partielle du méthane est, par contre, exothermique et offre donc l'avantage d'un procédé autotherme. De plus, la réaction d'OPM permet d'obtenir un gaz de synthèse présentant un rapport $H_2/CO$ égal à 2, ce qui est très intéressant dans le cadre d'une utilisation directe pour les réactions de Fischer-Tropsh. Elle peut également être envisagée pour la production d'hydrogène et trouve donc une application directe pour l'alimentation des piles à combustible.

L'objectif de cette étude est d'élucider le mécanisme de l'OPM en présence d'oxyde de lanthane $La_2O_3$ comme catalyseur car la connaissance du mécanisme réactionnel est primordiale pour le dimensionnement des futurs réacteurs et leur optimisation. L'étude expérimentale utilise un réacteur auto-agité catalytique qui comporte un volume gazeux dont l'agitation est assurée par des jets turbulents de réactif préchauffé. La réaction a été étudiée sur une large gamme de températures (650 à 850°C) avec des quantités de catalyseur variables.

La réaction étant réalisée à haute température, la présence de réactions radicalaires en phase gazeuse est inéluctable et ces réactions sont couplées aux réactions catalytiques. Le mécanisme a été élucidé en confrontant les résultats expérimentaux à ceux obtenus par simulation d'un mécanisme hétéro-homogène grâce au logiciel de simulation de mécanismes réactionnels Chemkin-Surface.

## 2. Technique expérimentale.

Le réacteur RAC qui a déjà été utilisé pour l'étude du couplage oxydant du methane (Simon et al., 2007) est constitué d'une partie hémisphérique de 60 mm de diamètre contenant la croix d'injection et prolongée par un cylindre. Un piston amovible à surface plane permet de déposer le catalyseur. Les deux parties peuvent s'emboîter offrant un volume de réacteur fixe égal à 110 cm$^3$.

---

[*] Auteur à qui la correspondance devrait être adressée : Paul-Marie.Marquaire@ensic.inpl-nancy.fr

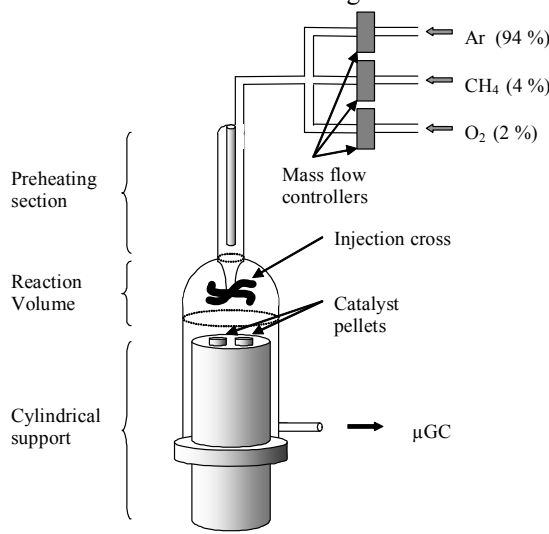

*Figure 1. Schéma du réacteur auto-agité catalytique*

Le débit des réactifs en entrée du réacteur est contrôlé par des régulateurs de débits massiques (RDM 280 de la société Air Liquide Alphagaz).

Avant l'entrée dans la demi-sphère, les gaz sont préchauffés et après, ils sont refroidis par une trempe à l'air. Le réacteur est en quartz car le chauffage du réacteur peut atteindre 900°C.

Les critères et règles de construction sont celles définies initialement par (David et Matras, 1975) pour un réacteur sphérique et s'appliquent également pour le réacteur RAC (David et al., 1979).

Le chauffage du réacteur est assuré par des fils chauffant de 2mm (thermocoax) de diamètre enroulés autour du réacteur au contact de la paroi.

La réaction se déroule dans la partie hémisphérique dont le volume est environ égal à 110 cm$^3$. L'avancement de la réaction dans la partie annulaire du préchauffage est négligeable car ce dernier ne représente que 1% du volume du réacteur.

L'analyse des gaz se fait en ligne à l'aide d'un micro chromatographe Agilent 3000 à 2 voies. Il est composé de deux colonnes, l'une étant une colonne tamis moléculaire 5A et l'autre étant une colonne capillaire PLOT U (*Porous Layer Open Tubular*). Les deux détecteurs associés à ces colonnes sont des catharomètres TCD.

Le système permet de détecter 8 produits en moins de 3 minutes. L'hydrogène $H_2$, l'oxygène $O_2$, le méthane $CH_4$ et le monoxyde de carbone CO sont détectés dans cet ordre sur la colonne tamis moléculaire tandis que le dioxyde de carbone $CO_2$, l'éthylène $C_2H_4$, l'éthane $C_2H_6$ et l'acétylène $C_2H_2$ sont détectés dans cet ordre sur la colonne Plot U.

Dans le réacteur RAC, les catalyseurs ne sont pas introduits sous forme de poudre mais sous forme de pastilles compactes obtenues à l'aide d'une matrice de type Macro-Micro KBr Die fournie par Aldrich. Pour ce faire, 0,4 g de poudre est compactée à une pression de 1,6 kbar (soit 20 kN) pour obtenir une pastille de diamètre 12,6 mm et de 1 mm d'épaisseur.

La composition d'entrée des gaz dans le réacteur est telle que $CH_4 / O_2 / Ar = 4 / 2 / 94$. Le rapport $CH_4/O_2$ est égal à 2 car c'est le rapport optimal pour conduire la réaction d'OPM dans nos conditions. De plus, les réactifs sont largement dilués dans l'inerte (Argon) afin non seulement d'éviter les variations de débits volumiques dues à l'avancement de la réaction mais aussi afin de mieux contrôler la température de la réaction et de limiter ainsi la présence de points chauds.

L'étude de la réaction de l'oxydation partielle du méthane (OPM) a été réalisée en faisant varier divers paramètres :

- la température de réaction T : comprises entre 500 et 900°C
- le temps de passage $\tau$ : compris entre 0,6 et 6 s.
- le temps de contact $t_c$ : compris entre 0 et 45 mg.s.cm$^{-3}$

## 3. Modélisation des résultats

### 3.1. Le mécanisme homogène

Nous avons utilisé le mécanisme écrit au DCPR et connu sous le nom «Base $C_0$-$C_2$» (Barbé et al., 1995 ; Fournet et al., 1999). Ce mécanisme homogène se compose de 444 réactions élémentaires réversibles ou directes pour lesquelles seules les espèces contenant moins de 3 atomes de carbone sont considérées.

Les paramètres cinétiques proviennent essentiellement de la littérature. Ces paramètres sont ceux de l'équation d'Arrhénius généralisée dont l'expression générale est :

$$k = A T^n \exp(-E_a/RT) \quad A\,(mol, cm^3, s) \quad E_a\,(cal.mol^{-1})$$

La première étape dans le développement d'un mécanisme complet est de s'assurer que le mécanisme homogène peut rendre compte des observations expérimentales. Aussi, des expériences sans catalyseur ont été réalisées à 810°C et 875°C en fonction du temps de passage. La comparaison entre résultats expérimentaux et simulations est présentée sur la Figure 2.

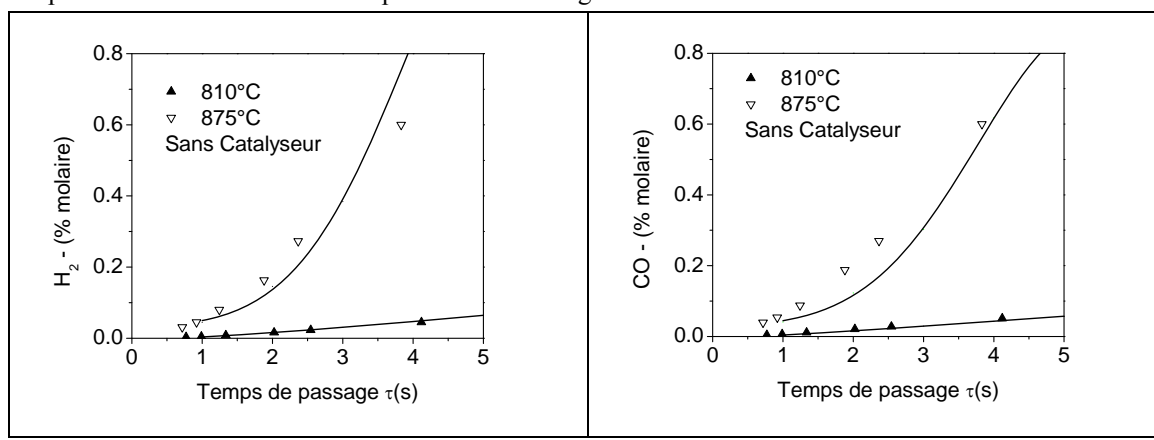

*Figure 2. Concentration en $H_2$ et CO en fonction du temps de passage à 810°C et 875°C sans catalyseur.*
*Comparaison entre expérience (symbole) et simulation (trait).*

Les produits majeurs sont $H_2$, CO et $CO_2$ tandis que les produits mineurs, voire les traces, sont $C_2H_4$ et $C_2H_6$. Notons que $C_2H_2$ n'est pas détecté dans les conditions opératoires utilisées. La concentration des produits majoritaires est environ 10 fois supérieure à celle des produits minoritaires lorsque le temps de passage est égal à 3s.

Les concentrations de sortie des différents composés augmentent avec le temps de passage et la température. Notons que cette augmentation est moins rapide pour $C_2H_6$ car l'éthane peut se décomposer en $C_2H_4$ et en produits oxygénés dans ces conditions opératoires

La Figure 2 montre qu'il y a un bon accord entre les données expérimentales et les valeurs simulées. Le mécanisme homogène « $C_0$-$C_2$ » permet de reproduire de manière satisfaisante les résultats expérimentaux. Par conséquent, c'est ce mécanisme qui sera utilisé par la suite.

### 3.2. Le mécanisme hétérogène

La nature exacte des sites actifs de l'oxyde de lanthane est mal connue. Différentes études suggèrent que les ions $O^{2-}$ sont susceptibles d'être les sites actifs ou d'être les précurseurs des espèces oxygénées pour l'activation du méthane (Yang et al., 1994 ; Hutchings et al., 1989a et b) ont proposé l'idée que les ions $O^-$ sont les espèces responsables de la forte production de radicaux méthyles tandis que les ions $O_2^{2-}$ sont responsables de la formation de radicaux $CH_2^{\cdot}$.

*Tableau 1. Mécanisme hétérogène complet écrit pour l'oxydation partielle du méthane. z représente le nombre de sites adjacents.*

| N° | Réactions de Surface | A (mol, cm$^3$, s) | n | E$_a$ (cal. mol$^{-1}$) |
|---|---|---|---|---|
| 1a | $O_2 + s \rightarrow O_2(s)$ | 3,0E6 | 0,0 | 1,5E3 |
| -1a | $O_2(s) \rightarrow O_2 + s$ | 2,3E13 | 0,0 | 45,0E3 |
| 2a | $O_2(s) + s \rightarrow 2 O(s)$ | 5,3E23z | 0,0 | 25,0E3 |
| -2a | $2 O(s) \rightarrow O_2(s) + s$ | 1,3E23z | 0,0 | 33,0E3 |
| 3a | $CH_4 + O(s) \rightarrow CH_3 + OH(s)$ | 3,0E8 | 0,0 | 8,84E3 |
| 4a | $CH_4 + s \rightarrow CH_3 + H(s)$ | 9,49E7 | 0,0 | 9,85E3 |
| 5a | $C_2H_6 + O(s) \rightarrow C_2H_5 + OH(s)$ | 8,7E9 | 0,0 | 3,0E3 |
| 6a | $C_2H_6 + s \rightarrow H(s) + C_2H_5$ | 9,8E7 | 0,0 | 5,0E3 |
| 7a | $C_2H_4 + O(s) \rightarrow C_2H_3 + OH(s)$ | 1,1E10 | 0,0 | 3,0E3 |
| 8a | $C_2H_4 + s \rightarrow H(s) + C_2H_3$ | 6,9E7 | 0,0 | 5,0E3 |
| 9a | $C_2H_5 + O(s) \rightarrow C_2H_4 + OH(s)$ | 5,5E7 | 0,0 | 0,0 |
| 10a | $C_3H_7 + O(s) \rightarrow C_3H_6 + OH(s)$ | 6,1E7 | 0,0 | 0,0 |
| 11a | $CH_3 + O(s) \rightarrow CH_2 + OH(s)$ | 1,9E9 | 0,0 | 2,8E3 |
| 12a | $CH_2 + O(s) \rightarrow CH + OH(s)$ | 3,6E11 | 0,0 | 11,9E3 |
| 13a | $CH + O(s) \rightarrow C + OH(s)$ | 8,9E8 | 0,0 | 4,7E3 |
| 14a | $CH_3 + O(s) \rightarrow CH_3O(s)$ | 9,9E8 | 0,0 | 0,6E3 |
| 15a | $CH_3O(s) + O(s) \rightarrow HCHO + OH(s) + s$ | 1,2E23z | 0,0 | 0,0 |
| 16a | $HCHO + O(s) \rightarrow CHO + OH(s)$ | 3,4E7 | 0,0 | 3,0E3 |
| 17a | $CHO + O(s) \rightarrow CO + OH(s)$ | 6,9E7 | 0,0 | 0,0 |
| 18a | $C_2H_5 + O(s) \rightarrow C_2H_5O(s)$ | 1,0E9 | 0,0 | 0,6E3 |
| 19a | $C_2H_5O(s) + O(s) \rightarrow CH_3CHO + OH(s) + s$ | 6,6E21z | 0,0 | 0,0 |
| 20a | $CO + O(s) \rightarrow CO_2 + (s)$ | 8,31E8 | 0,0 | 0,0 |
| 21a | $CO_2 + (s) \rightarrow CO_2(s)$ | 6,2E8 | 0,0 | 0,0 |
| -21a | $CO_2(s) \rightarrow CO_2 + (s)$ | 2,3E13 | 0,0 | 43,58E3 |
| 22a | $C + O(s) \rightarrow CO(s)$ | 1,1E11 | 0,0 | 0,0 |
| 23a | $C + O(s) \rightarrow CO + s$ | 1,1E11 | 0,0 | 0,0 |
| 24a | $CO(s) + O(s) \rightarrow CO_2(s) + s$ | 1,1E23z | 0,0 | 0,0 |
| 25a | $H + s \rightarrow H(s)$ | 2,3E13 | 0,0 | 0,0 |
| 26a | $H(s) + H(s) \rightarrow H_2 + 2 s$ | 4,0E23z | 0,0 | 0,0 |
| 27a | $OH(s) + H(s) \rightarrow H_2O + 2 s$ | 1,0E22z | 0,0 | 0,0 |
| 28a | $OH(s) + OH(s) \rightarrow H_2O + O(s) + s$ | 3,0E23z | 0,0 | 2,4E3 |
| 29a | $H_2 + s \rightarrow 2 H(s)$ | 6,1E16 | 0,0 | 0,0 |
| 30a | $H_2 + O(s) \rightarrow OH(s) + H$ | 1,0E9 | 0,0 | 0,0 |

(Lacombe et al., 1994) ont identifié différents sites actifs. D'abord, les sites basiques associés à des lacunes d'oxygène qui engendreraient la dissociation de l'oxygène gazeux en atomes d'oxygène adsorbés à la surface du catalyseur. Ces atomes seraient ensuite capables d'activer les molécules de méthane. Ensuite, les atomes peu coordonnés sur lesquels les radicaux méthyles réagiraient pour donner des composés oxydés, tels que $CO_2$. La conclusion de ces différents auteurs est qu'il existe au moins deux types de sites actifs ou deux espèces actives pour la réaction d'oxydation partielle du méthane.

C'est ainsi que dans le mécanisme hétérogène proposé, deux sites ont été considérés. Ces sites sont notés s et O(s). Le premier fait référence à un site non identifié précisément, ce pourrait être par exemple un atome de lanthane. Tandis que le second, O(s), fait référence à l'oxydation du site s.

Dans notre mécanisme, l'oxygène gazeux se dissocie en deux atomes d'oxygène adsorbés actifs, en accord avec de précédentes études sur la chimisorption de l'oxygène (Huang et al., 2000). L'interaction entre l'oxygène gazeux et les atomes de la structure cristallographique de $La_2O_3$ est supposée rapide (Lacombe et al., 1995).

Le mécanisme est ensuite écrit de manière systématique en considérant les différentes réactions possibles entre les produits formés ou les radicaux les plus importants et les sites s et O(s). Les réactions sont écrites selon le formalisme de Eley-Rideal dans lequel une molécule en phase gazeuse réagit avec un site actif. Ainsi, dans le cas de l'amorçage du méthane, les deux réactions considérées sont les *réactions 3a et*

*4a* (voir Tableau 1). La réaction du méthane avec deux sites s et O(s) a déjà été considérée avec succès dans la littérature par (Toops et al., 2002). Elle a été également envisagée par (Amorebieta et al., 1989) dans le cas de l'oxydation du méthane avec de l'oxyde de samarium. Nous supposons à travers la *réaction 3a* que c'est également le cas pour l'oxyde de lanthane $La_2O_3$. Il est intéressant de remarquer que ces deux réactions produisent des radicaux méthyles ainsi que les espèces de surface OH(s) et H(s).

De la même façon, les réactions de $C_2H_6$ et $C_2H_4$ avec les sites s et O(s) forment les radicaux $C_2H_5\cdot$ et $C_2H_3\cdot$ dans la phase gazeuse, ainsi que OH(s) et H(s). Les réactions entre H(s) et OH(s) sont à l'origine de la formation d'hydrogène et d'eau à la surface du catalyseur selon les *réactions 26a, 27a et 28a* (voir Tableau 1).

Le mécanisme hétérogène est composé de 33 réactions élémentaires. Les facteurs préexponentiels sont calculés à partir des fonctions de partition des réactifs et du complexe active par des méthodes dérivées des techniques de Benson (Benson, 1976) tandis que les énergies d'activation ont été choisies en première approximation par analogie avec des réactions en phase gazeuse.

### 3.3. Validation du mécanisme hétéro-homogène

Afin de valider ou d'infirmer le mécanisme proposé, des simulations ont été effectuées en utilisant le mécanisme dans sa totalité, c'est-à-dire en prenant en compte simultanément le mécanisme homogène et le mécanisme hétérogène. Dans le cas du mécanisme hétérogène, les valeurs des facteurs préexponentiels ont éventuellement été modifiées par un coefficient multiplicateur qui reste dans l'ordre de grandeur de la précision des estimations. Les valeurs données dans le Tableau 1 sont les valeurs obtenues après ajustement.

Afin de s'assurer que le mécanisme rend compte de manière satisfaisante des observations expérimentales pour la réaction de l'oxydation partielle du méthane, la comparaison entre simulations et expériences a été réalisée sur un large domaine en faisant varier de nombreux paramètres. Classiquement, les comparaisons ont été d'abord réalisées en faisant varier le temps de passage à deux températures différentes (700°C et 850°C) et pour une quantité donnée de catalyseur (1 pastille de $La_2O_3$ de 0,4 g). Les résultats sont donnés sur la Figure 3.

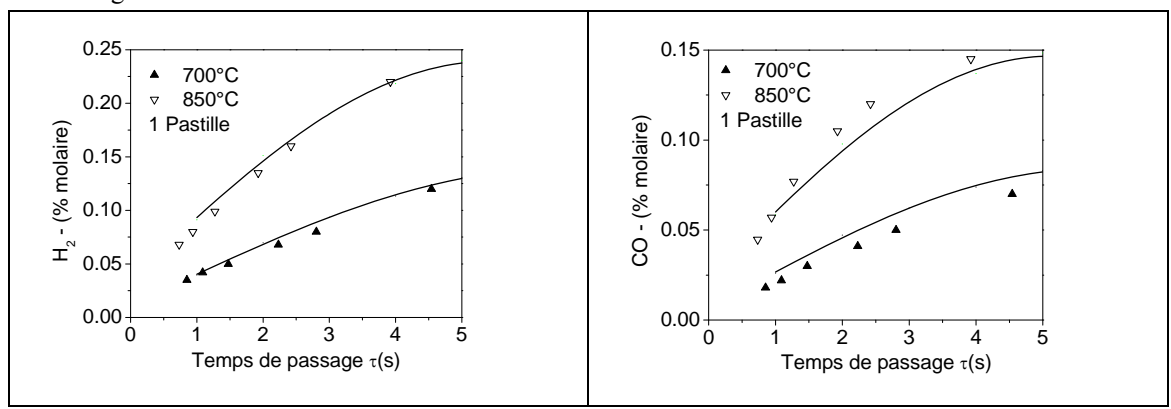

*Figure 3. Concentration en H2 et CO en fonction du temps de passage à 700°C et 850°C avec 1 pastille de $La_2O_3$. Comparaison entre expérience (symbole) et simulation (trait).*

La comparaison entre valeurs simulées et valeurs expérimentales est satisfaisante : les tendances générales sont en accord et il en va de même pour les valeurs des concentrations de sortie.

Par conséquent, nous pouvons considérer que le mécanisme proposé est un mécanisme adéquat pour la modélisation de l'oxydation partielle du méthane avec de l'oxyde de lanthane $La_2O_3$.

## 4. Analyse et discussion du mécanisme proposé

Un schéma mécanistique général est fourni sur la Figure 4. Sur cette figure, les flèches grisées sont associées à des réactions hétérogènes tandis que les flèches noires sont pour les réactions homogènes.

Le méthane peut être activé et décomposé en radicaux méthyles par l'intermédiaire de réactions hétérogènes (étape 1) ou de réactions homogènes (étape 2). Dans ce dernier cas, R˙ désigne un ensemble de radicaux pouvant être OH˙, H˙, $HO_2$˙ ou O˙˙. Dans l'étape 3, l'hydrogène est consommé par réaction avec les radicaux méthyles pour redonner du méthane et de l'hydrogène adsorbé H(s). Les radicaux méthyles ont un rôle central dans le mécanisme. Ils peuvent suivre deux voies de conversion distinctes. La première est la *voie pyrolytique* (étapes 4 à 11). Elle conduit à la formation de composés exempts d'oxygène à savoir essentiellement les composés $C_2$. La seconde est la *voie oxydante* (étapes 12 à 21), elle conduit à la formation de radicaux et de composés oxygénés tels que le formaldéhyde HCHO, le monoxyde de carbone CO et le dioxyde de carbone $CO_2$. Ces deux voies de conversion sont connectées par l'oxydation possible de radicaux $C_2$ et de l'acétylène en radicaux $HO_2$˙, CHO˙ et en formaldéhyde (étapes 22 à 24).

Notons enfin qu'une troisième voie réactionnelle est indiquée sur la Figure 4. Il s'agit d'une succession d'étapes hétérogènes conduisant à la formation d'hydrogène et d'eau par réaction entre groupes H(s) et OH(s) adsorbés à la surface du catalyseur. L'origine des groupes H(s) provient essentiellement de l'adsorption de radicaux H˙ tandis que les groupes OH(s) sont produits majoritairement par des réactions types Eley-Rideal entre O(s) et un hydrocarbure, tels que méthane, éthane ou éthylène.

Il est important de noter que les radicaux méthyles peuvent réagir avec l'hydrogène pour former des radicaux H˙ et du méthane. Les radicaux H˙ s'adsorbent majoritairement sur la surface du catalyseur en H(s). La réaction entre hydrogène adsorbé H(s) reforme du dihydrogène $H_2$. Il existe donc un cycle interne pour l'hydrogène ce qui suggère que la conversion du méthane ainsi que la sélectivité de l'hydrogène seront limitées à fort avancement. C'est exactement ce qui est observée expérimentalement.

Notre mécanisme montre que pour de faibles conversions, les radicaux H˙ réagissent avec le méthane pour former des radicaux méthyles et de l'hydrogène tandis qu'à plus forte conversion, lorsque $XCH_4 > 0,1$, la réaction se déplace en sens inverse et c'est alors la réaction entre radicaux méthyles et hydrogène qui a lieu. Ceci est dû notamment à la concentration en radicaux méthyles qui augmente avec la conversion. La réaction $CH_4 + H\cdot = CH_3\cdot + H_2$ peut être considérée comme une réaction tampon (*Buffer reaction*) précédemment décrite par (Weissman et Benson, 1984). Par exemple, à 700°C avec 1 pastille, l'étape 3 n'a pas lieu. Au contraire dans ces conditions c'est la réaction entre méthane et radicaux H˙ qui prédomine et cette réaction est inclue dans l'étape 2. Par contre, à 850°C avec 8 pastilles. 42% des radicaux méthyles sont consommés pour reformer du méthane, ce qui est tout à fait considérable.

Les voies pyrolytique et oxydante sont connectées par les étapes 22, 23 et 24. Ces étapes mettent en évidence le fait que les composés $C_2$ peuvent être oxydés ce qui implique que le rendement en $C_2$ peut atteindre une valeur limite (Mc Carty et al., 1989 ; Mc Carty et al., 1991). C'est une observation problématique dans le cas de la réaction du couplage oxydant du méthane dont l'objectif premier est la formation de $C_2$. Ceci est en accord avec (Lin et al., 1986) qui ont montré que l'oxyde de lanthane $La_2O_3$ est actif pour la dégradation de l'éthane $C_2H_6$. Ainsi, les auteurs ont pu obtenir une sélectivité en $C_2$ égale à 47% à faible conversion de méthane seulement (9,4%).

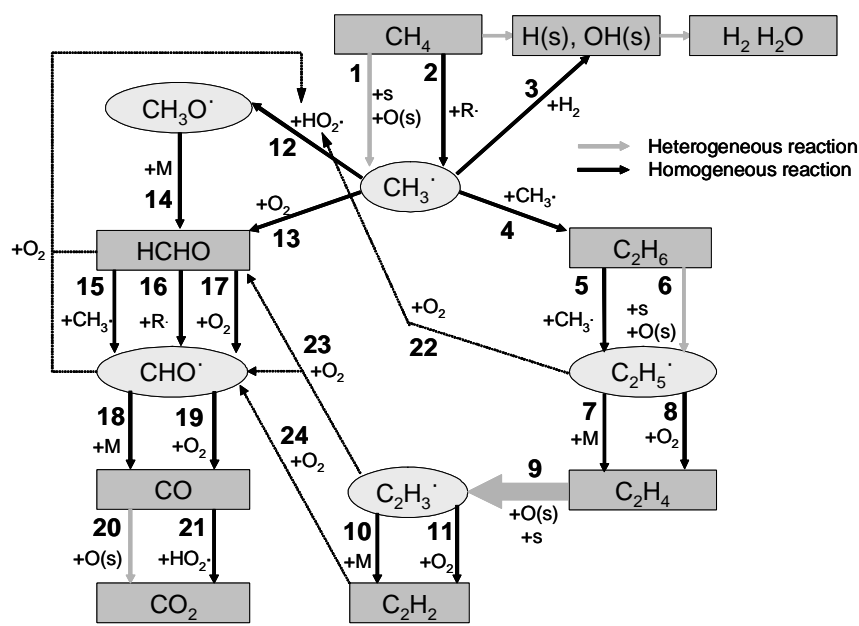

*Figure 4. Schéma mécanistique général pour la réaction hétéro-homogène de l'oxydation partielle du méthane avec $La_2O_3$.*

L'acétylène est presque totalement converti en produits oxygénés quelques soient les conditions opératoires. C'est pourquoi, la quantité d'acétylène détectée expérimentalement est très faible, à des niveaux de traces. Ainsi, la formation de $CO_x$ (x=1 ou 2) provient non seulement de la voie oxydante mais également de l'oxydation des composés $C_2$. Ceci est en accord avec (Ekstrom et al., 1989) qui ont montré que les oxydes de carbone sont formés de manière importante par l'oxydation des produits de la réaction, tels que les produits $C_2$, lorsque des catalyseurs à base d'oxydes de terres rares sont utilisés.

## 5. Conclusion

Le mécanisme proposé rend bien compte des observations expérimentales obtenues dans des conditions opératoires favorables à l'oxydation partielle du méthane. Il est alors possible d'effectuer une analyse détaillée du mécanisme en mettant en évidence la nature et l'importance des différentes étapes intervenant dans le mécanisme global. Ces étapes peuvent être homogènes ou hétérogènes et plus ou moins significatives selon la valeur des paramètres opératoires. Elles traduisent le comportement et le rôle du catalyseur. Dans notre cas, elles confirment que le rôle premier du catalyseur est de former des radicaux qui vont réagir en phase gazeuse. L'ensemble des réactions entre en compétition et l'analyse quantitative dans diverses conditions nous a permis de comprendre l'origine du faible rendement en hydrogène pour de fortes conversions de méthane. Nous avons également pu observer des similitudes avec le mécanisme du couplage oxydant du méthane (OCM) dans lequel les produits $C_2$ ont un rendement limité.

**Références**

Amorebieta, V.T.; Colussi, A.J., 1989. Mass Spectrometric Studies of the Low-Pressure Oxidation of Methane on Samarium Sesquioxide. J. Phys. Chem. 93, 5155-5158.

Barbe, P.; Battin-Leclerc, F.; Côme, G.M., 1995. Experimental and modelling study of methane and ethane oxidation between 773 and 1573 K. J. Chim. Phys. 92, 1666-1692.

Benson, S.W., 1976. Thermochemical Kinetics: Methods for the Estimation of Thermochemical Data and Rate Parameters. 2nd Edition Wiley Interscience

David, R.; Matras, D., Rules for construction and extrapolation of reactors self-stirred by gas jets, 1975. Can. J. Chem. Eng. 53(3), 297-300.

David, R.; Houzelot, J.L.; Villermaux, J., 1979. Gas mixing in jet-stirred reactors with short residence times. Proceedings of the European Conference on Mixing $3^{rd}$ 1, 113-124.

Ekstrom, A.; Lapszewicz, J.A., 1989. A Study of the Mechanism of the Partial Oxidation of Methane over Rare Earth Oxide Catalysts Using Isotope Transient Techniques. J. Phys. Chem. 93, 5230-5237.

Fournet, R.; Bauge, J.C.; Battin-Leclerc, F., 1999. Experimental and Modeling of Oxidation of Acetylene, Propyne, Allene and 1,3-Butadiene. Int. J. Chem. Kin. 31, 361-379.

Huang, S.J.; Walters, A.B.; Vannice, M.A., 2000. Adsorption and Decomposition of NO on Lanthanum Oxide. J. Catal. 192, 29-47.

Hutchings, G.J.; Woodhouse, J.R.; Scurrell, M.S., 1989a. Partial Oxidation of Methane over Oxide Catalysts. Comments on the Reaction Mechanism. J. Chem. Soc., Faraday Trans. 85(8), 2507-2523.

Hutchings, G.J.; Scurrell, M.S.; Woodhouse, J.R., 1989b. Partial Oxidation of Methane over Samarium and lanthanum Oxides: a study of the reaction mechanism. Cat. Today 4, 371-381.

Lacombe, S.; Geantet, C.; Mirodatos, C., 1994. Oxidative Coupling of Methane over Lanthana Catalysts. I. Identification and Role of Specific Active Sites. J. Catal. 151, 439-452.

Lacombe, S.; Zanthoff, H.; Mirodatos, C., 1995. Oxidative Coupling of Methane over Lanthana Catalysts. II. A Mechanistic Study Using Isotope Transient Kinetics, J. Catal. 155, 106-116.

Lin, C.-H.; Campbell, K.D.; Wang, J.-X.; Lunsford, J.H., 1986. Oxidative Dimerization of Methane over Lanthanum Oxide, J. Phys. Chem. 90, 534-537.

Mc Carty, J.G.; Mc Ewen, A.B.; Quinlan, M.A., 1989. Inherent limitation in the direct catalytic conversion of natural gas into higher hydrocarbons. International Gas Research Conference, Tokyo, 6-9 Nov.

Simon, Y.; F. Baronnet; P.M. Marquaire, 2007. Kinetic modelling of the oxidative coupling of methane. Ind. Eng. Chem. Res. (sous presse).

Toops, T.J., Walters, A.B., Vannice, M.A., 2002. Methane combustion over $La_2O_3$-based catalysts and $\gamma$-$Al_2O_3$. Appl. Catal. A 233, 125-140.

Weissman, M.; Benson, S.W., 1984. Pyrolysis of Methyl Chloride, a Pathway in the Chlorine-Catalyzed polymerization of Methane. Int. J. Chem. Kin. 16, 307-333.

Yang, T.; Feng, L.; Shen, S., 1994. Oxygen Species on the Surface of $La_2O_3$/CaO and its Role in the Oxidative Coupling of Methane, J. Cat. 145, 384-389.